\documentclass[referee]{aa}
\usepackage{txfonts}
\usepackage{graphicx}
\usepackage[final]{epsfig}
\usepackage{psfrag}
\begin{document}
   \title{Highly accurate calculation of rotating neutron stars}
 
   \subtitle{Detailed description of the numerical methods}

   \author{M. Ansorg, A. Kleinw\"achter, and R. Meinel
          }
   \authorrunning{Ansorg et al.}
   \titlerunning{Highly accurate calculation of rotating neutron stars}

   \offprints{M. Ansorg,\\ \email{ansorg@tpi.uni-jena.de}}

   \institute{Theoretisch-Physikalisches Institut, University of Jena,
              Max-Wien-Platz 1, 07743 Jena, Germany\\
              }

   \date{Received / Accepted  }

   \abstract{
   We give a detailed description of the recently developed multi-domain 
   spectral method for constructing highly accurate general-relativistic 
   models of rapidly rotating stars. For both `ordinary' and
   `critical' configurations, it is exhibited by means of
   representative examples, how the accuracy improves as the
   order of the approximation increases. Apart from homogeneous fluid
   bodies, we also discuss models of polytropic and strange stars. 
   \keywords{stars: rotation ---
                stars: neutron ---
                gravitation ---
                relativity ---
                methods: numerical
               }
               }
   \maketitle
%
\section{Introduction}

The structure and the gravitational field of
relativistic, axisymmetric and stationary, uniformly
rotating perfect fluid bodies is investigated in order to model extraordinarily 
compact astrophysical objects like neutron stars. The numerical calculation 
of these objects was the subject of papers by several authors, see
Bonazzola \& Schneider \cite{bs}, Wilson \cite{wi},
Butterworth \& Ipser \cite{butips1, butips2}, Friedman et al. \cite{fip1,fip2},
Lattimer et al. \cite{latt}, Neugebauer \& Herold \cite{nh}, Herold \&
Neugebauer \cite{hn}, Komatsu et al. \cite{keh1, keh2}, Eriguchi et al.
\cite{ehn}, Stergioulas \& Friedman \cite{ster}, Bonazzola et al. \cite{bgsm};
for reviews see Friedman \cite{fri} and Stergioulas \cite{sterg}. 

The idea to use a `multi-domain 
spectral method' was introduced by Bonazzola et al. (\cite{bgsm,bgm}).
In their `BGSM-code', the space of physical coordinates is divided into
several subregions, each one of them to be mapped onto a 
cross product of intervals. The physical field
quantities are expressed in a spectral expansion with respect to all
coordinates defined on the specific intervals. 
If the interior region of the star is
chosen to be exactly one of the domains, then it is possible to obtain
representations of the field quantities that are smooth functions 
within the cross product of intervals. 
The spectral expansions then provide a 
very precise approximation of the field quantities.
In particular, this choice for the domains circumvents 
the occurrence of the
Gibbs phenomenon at the star's surface, which appears when non-smooth 
physical fields (such as the mass-energy density) are 
expressed in terms of a spectral expansion.

The subject of this paper is a detailed description of our  
multi-domain spectral method (hereafter `AKM-method') 
which led to an accuracy of up to 12
digits for rapidly rotating, strongly relativistic 
homogeneous fluid bodies (Ansorg, Kleinw\"{a}chter, \&
	  Meinel \cite{AKM02}).
Its further development enabled us to calculate
rotating toroidal bodies (the relativistic Dyson rings; see Ansorg, Kleinw\"{a}chter, \&
	  Meinel \cite{AKM03a}).

The corresponding fundamental features of the AKM-method are as described 
above for the BGSM-code. However, the methods differ in the following:
\begin{enumerate}
  \item Instead of the three domains in the BGSM-code (two of them 
  exterior to the star), we have only two domains, since we do not split up the
  region exterior to the star.
  \item As described in Bonazzola et al. (\cite{bgm}), the BGSM-code is an iterative scheme
  with each iteration step consisting of several procedures
  including the solving of nonlinear Poisson-like equations and
  the determination of an improved approximation of the star's surface.
  In the AKM-method, a large set of nonlinear algebraic equations
  for the unknown spectral coefficients corresponding to all field
  quantities and the unknown shape of the star's surface is
  simultaneously solved by a Newton-Raphson method. 
  \item For the AKM-method, the restriction to only two domains, 
  one of them exactly corresponding to the interior region of the star, 
  can be maintained when the mass shedding limit is approached. 
  In this limit, the star is characterized by a cusp at its equatorial rim, 
  thus requiring that the inner-most domain of the BGSM-code deviate slightly
  from the star's interior (see Bonazzola et al. \cite{bgm}; the
  resulting Gibbs phenomenon is limited since the displacement is
  small). 
\end{enumerate}
We begin our description of the AKM-method 
with a review of the line element and
Einstein's field equations, boundary, regularity and transition
conditions as well as the resulting free boundary value problem. 
Following this, we provide an introduction to the method in question
containing a comprehensive description of the specific spectral field
representations and the resulting set of nonlinear algebraic equations 
that ensures both the validity of the field
equations within each domain and the transition conditions at the
domains' boundary. In the subsequent section, we apply the AKM-method
to `ordinary' homogeneous, strange as well as polytropic
stars (with a polytropic exponent $\Gamma=2$) and display the
improvement of the accuracy as the order of the approximation
increases. Here, `ordinary' stars means configurations that are not
special in the sense that they neither rotate at the mass shedding
limit, possess infinite central
pressure, nor show a considerable oblateness, but may be strongly
relativistic as in the example given in table 11 of Nozawa et al. (\cite{nozawa}) 
and table 1 of Ansorg et al. (\cite{AKM02}).

Then we give representative examples concerning the 
excluded `critical' configurations. At first, we consider
homogeneous, strange as well as polytropic stars at the mass shedding limit.
Following this, we study homogeneous configurations with an infinite 
central pressure. Finally, the last part of this section is devoted to
highly flattened homogeneous bodies. 

In what follows, units are used in which the speed of light as
well as Newton's constant of gravitation are equal to 1.
 
\section{Line element, Einstein's field equations, and the free
boundary value problem}
We use two different formulations of the line element describing the
gravitational field of a uniformly rotating 
perfect fluid body\footnote{The basic principles of the AKM-method rely on a
rapidly converging Chebyshev representation of all physical and geometrical
quantities within appropriate coordinates. The method is therefore applicable 
to arbitrary differentially rotating configurations (with some analytical
rotation law). 
In this article we restrict ourselves to uniform rotation. In
this case we may use the simpler field equations valid within the comoving frame
of reference.}.
The corresponding Lewis-Papapetrou coordinates $(\rho,\zeta,\varphi,t)$ are
uniquely defined 
by the requirement that the metric coefficients and their first derivatives 
be continuous at the surface of the body. 

Exterior to the star in question we write:
\begin{eqnarray}ds^2={\rm e}^{2\alpha}(d\rho^2+d\zeta^2)
       +W^2{\rm e}^{-2\nu}(d\varphi-\omega\, dt)^2
        -{\rm e}^{2\nu}dt^2\end{eqnarray}
while for the interior we take metric functions valid in the comoving
frame of reference:
\begin{eqnarray}ds^2={\rm e}^{-2U'}[{\rm e}^{2k'}(d\rho^2+d\zeta^2)+W^2 d\varphi'^2]
         -{\rm e}^{2U'}(dt+a'd\varphi')^2\end{eqnarray}
Here, the only new coordinate is $\varphi'=\varphi-\Omega t$ where
$\Omega$ is the (constant) angular velocity of the star.

Hence, we get the following transformation formulae:
\begin{eqnarray}\label{Transf}
  W^{-1}{\rm e}^{2\nu}\pm(\omega-\Omega)
    =\left(W{\rm e}^{-2 U'}\mp a'\right)^{-1}, \quad \alpha=k'-U'\end{eqnarray}
The exterior field equations following from the above line element
read as follows\footnote{Once $\nu,\omega$ and $W$ are known, the potential
$\alpha$ can be determined by a line integral 
(with $\alpha\to 0$ as $\rho^2+\zeta^2\to\infty$).}:
\begin{eqnarray} \label{Exterior}
     \triangle_W^{(1)}\nu &=& \frac{1}{2}W^2{\rm e}^{-4\nu}
                             (\omega_\rho^2+\omega_\zeta^2) \\[3mm]
     \triangle_W^{(3)}\omega &=& 4 (\nu_\rho\omega_\rho+\nu_\zeta\omega_\zeta) \\[3mm]
     \triangle_2   W          &=& 0 
\end{eqnarray} 
with the abbreviations 
\begin{eqnarray} 
     \triangle_2       &=&\partial^2_\rho+\partial^2_\zeta\\[3mm]
     \triangle_W^{(j)} &=&\triangle_2+
                          j\,W^{-1}(W_\rho\partial_\rho
				  +W_\zeta\partial_\zeta).
\end{eqnarray} 
Since in the comoving frame the energy-momentum tensor reads
    \begin{eqnarray}T^{ik}=(\mu+p)u^iu^k+pg^{ik}\,,\quad
    u^k={\rm e}^{-U'}\delta_4^k\end{eqnarray}
where $\mu$ is the total mass-energy density and $p$ the pressure,
the interior field equations assume a particularly simple form\footnote{See,
for example, Kramer et al. (\cite{kra}).}:
\begin{eqnarray} \label{Interior}
     \triangle_W^{(1)}U'+\frac{1}{2}W^{-2}{\rm e}^{4U'} 
                             [(a'_\rho)^2+(a'_\zeta)^2]&=&
				     4\,\pi {\rm e}^{2(k'-U')}(\mu+3p)\\[3mm]
     \triangle_W^{(-1)}a' +4 (a'_\rho U'_\rho+a'_\zeta U'_\zeta) &=& 0\\[3mm]
     \triangle_2   W          &=& 16\,\pi {\rm e}^{2(k'-U')} \,W\,p 
\end{eqnarray} 
As with the above potential $\alpha$, the metric function $k'$ can be determined 
via a line integration from the potentials $U',a'$ and $W$ that follows from
\begin{eqnarray}  \label{k_prime}
   \begin{array}{l}
      \left(
	    \begin{array}{cc} 
	       W_\rho & -W_\zeta \\[3mm] W_\zeta & W_\rho
           \end{array}
	 \right)
       \left(
	    \begin{array}{l} 
	       k'_\rho \\[3mm] k'_\zeta
          \end{array}
	 \right)-
       \left(
	    \begin{array}{c} 
	       \frac{1}{2}(W_{\rho\rho}-W_{\zeta\zeta}) \\[3mm] W_{\rho\,\zeta}
          \end{array}
	 \right)=\\[1cm]\qquad=
       \left(
	    \begin{array}{c} 
             W[(U'_\rho)^2-(U'_\zeta)^2]-\frac{1}{4}W^{-1}{\rm e}^{4U'}
             [(a'_\rho)^2-(a'_\zeta)^2]\\[3mm]
             2W U'_\rho U'_\zeta -\frac{1}{2}W^{-1}{\rm e}^{4U'}
             a'_\rho a'_\zeta
          \end{array}
	 \right),
   \end{array}
\end{eqnarray} 
such that along the rotation axis $\rho=0$ the condition 
\begin{eqnarray} \label{k_axis}
{\rm e}^{k'}=\lim_{\rho\to 0} \frac{W(\rho,\zeta)}{\rho}
\end{eqnarray} 
holds. Additionally, for a given equation of state, $p=p(\mu) $ or
$\mu=\mu(p)$, the relativistic Euler equations $T^{ik}_{\;;k}=0$ yield
$\mu$ and $p$ in terms of the metric potential $U'$:
\begin{eqnarray} \label{Int_Eul}
{\rm e}^{U'}\exp\left[\int_0^p\frac{dp'}{\mu(p')+p'}\right]
={\rm e}^{V_0}=\mbox{const.}\end{eqnarray} 
Hence, for the exterior potentials $\nu,\omega,W$ as well as for the interior 
potentials $U',a',W$, particular systems of partial differential equations emerge. 
At the surface $B$ of the star, the pressure $p$ vanishes, which
leads to a constant surface potential $U'=V_0$, see Eq. (\ref{Int_Eul}).
If additionally the boundary $B$ as well as the corresponding
boundary values of the potentials $a'$ and $W$ were given, 
we would have to solve a particular interior
and exterior boundary value problem of the respective field
equations, completed by regularity conditions along the rotation axis
(here, the $\rho$-derivatives of $\nu,\omega,W/\rho,U'$ and $a'$ vanish)
and at infinity (here  $\nu\to 0$, $\omega\to 0$ and
$W-\rho\to 0$). However, we have to deal with
a free boundary value problem, where both the boundary $B$ and the 
values of $a'$ and $W$ at $B$ are unknown, but have
to be determined such that the normal derivatives of the potentials $U'$, $a'$
and $W$ behave continuously at $B$ \footnote{It is a consequence of
the field equations that $k'$ is then also differentiable.}. 

For a given
equation of state, the corresponding solution depends on two parameters,
e.g.~the angular velocity $\Omega$ and the gravitational mass $M$. Note that
there might be multiple solutions corresponding to a specific prescribed
parameter pair. For the description of the AKM-method, we consider at first the
particular prescription of the pair $(V_0,\Omega)$, 
but treat in a separate subsection the possible
prescription of other parameter pairs.

Together with the regularity conditions along the rotation axis, 
we assume that all metric potentials possess
reflectional symmetry with respect to the
equatorial plane $\zeta=0$,
leading to a vanishing $\zeta$-derivative in this plane 
(see, for example, Meinel \& Neugebauer \cite{meineu}).

\section{Description of the method}
\label{AKM}

A function, defined and analytic on a closed interval, 
can be represented by a rapidly converging Chebyshev-expansion. The
spirit of the AKM-method is to use this property for all 
gravitational potentials, boundary values, and the unknown shape of
the surface, which therefore need to be defined on appropriate
cross products of intervals. The corresponding
Chebyshev-coefficients are determined by a high-dimensional nonlinear 
set of algebraic equations that encompasses both field equations and
transition conditions and is solved by a Newton-Raphson method.

\subsection{The mappings of the subdomains}
\label{Mappings}

As already mentioned in the introduction, we divide the total space
of physical coordinates into two subregions, an inner domain covering
exactly the interior region of the star, and an outer one
encompassing the exterior vacuum region. Both subregions are mapped 
onto the square $I^2=[0,1]\times[0,1]$, which we realize by introducing  
a non-negative function $y_{\rm B}$ defined on the interval $I=[0,1]$ 
that describes the surface of the body by
\begin{eqnarray} \label{Surf_B}
\begin{array}{lcl}
  B=\{(\rho,\zeta)\mbox{:}\quad\rho^2=r_{\rm e}^2\hspace*{0.1mm}t,
                     \,\zeta^2=r_{\rm p}^2\hspace*{0.5mm}y_{\rm B}(t)\,,
                     \quad 0\leq t\leq 1\}\;, &&\\[5mm]
y_{\rm B}(0)=1\,,\quad y_{\rm B}(1)=0\;. &&\end{array}
\end{eqnarray}
Here $r_{\rm e}$ and $r_{\rm p}$ are the equatorial and polar coordinate radii of
the body respectively. 

A particular example for the mapping in question is given by 
\begin{eqnarray} \label{Transf_Int}
\rho^2=r_{\rm e}^2\hspace*{0.1mm}s\hspace*{0.1mm}t\,,\quad 
\zeta^2=r_{\rm p}^2\hspace*{0.1mm}s\hspace*{0.1mm}y_{\rm B}(t)\,,\quad (s,t)\in I^2
\end{eqnarray}
for the interior region and
\begin{eqnarray} \label{Transf_Ext}
\rho^2=\frac{r_{\rm e}^2\hspace*{0.1mm}t}{s^2}\,,\quad 
\zeta^2=\frac{r_{\rm p}^2\hspace*{0.5mm}y_{\rm B}(t)}{s^2}\,,\quad (s,t)\in I^2
\end{eqnarray}
for the region exterior to the star. In this manner, the axis $\rho=0$ and the plane $\zeta=0$ are mapped to the
coordinate boundaries $t=0$ and $t=1$ respectively. Furthermore, the surface
$B$ of the body is characterized by $s=1$. For the interior and exterior 
transformation, the point $s=0$ corresponds to the origin and 
to infinity respectively.

Writing $\rho^2$ and $\zeta^2$ (and not $\rho$ and $\zeta$)
in terms of the new variables $s$ and $t$ already takes the
regularity condition along the rotation axis as well as the reflectional
symmetry with respect to the equatorial plane into account. Indeed,
for any potential that is analytic with respect to the variables $s$ and
$t$, it follows that the $\rho$-derivative at $\rho=0$ as well as 
the $\zeta$-derivative at $\zeta=0$ vanishes provided the above coordinate
transformation is invertible there. 
The latter condition is only violated for $s=0$.\footnote{Note that the interior coordinate
transformation introduced in section \ref{Flat} is not invertible
at the equatorial rim of the star, but it is so at $s=0$.}

It turns out that the requirements of the regularity of the
potentials (as functions of $s$ and $t$) at $t=0$ and $t=1$ replace a
particular boundary condition here, that usually must be imposed.
Similarly, the regularity of the interior potentials  
supersedes a boundary condition at the coordinate's origin. 
However, the asymptotic
behaviour at infinity still must be considered, see section \ref{Repr_Fields}.

Note that for critical configurations we need to modify the above mapping, 
see sections \ref{Shed} and \ref{Flat}. 
\subsection{The representations of the potentials and the surface}
\label{Repr_Fields}
For each of the gravitational potentials we use a specific
Chebyshev-expansion that takes known boundary and transition
conditions into account. In particular we know ($r^2=\rho^2+\zeta^2$):
\begin{equation}
\label{MJ_out}
\lim\limits_{r\to \infty}(r \nu)=-M\;,\qquad \lim\limits_{r\to \infty}(r^3\omega)=2J\;,
\end{equation}
\begin{eqnarray}\begin{array}{lcl}
\left|\,\lim\limits_{r\to \infty}r^2\,(W\rho^{-1}-1)\,\right|&<&\infty\;,\\[4mm]
\left|\,\lim\limits_{\rho\to 0}a'\rho^{-2}\right|&<&\infty\;,\\[4mm]
\left|\,\lim\limits_{\rho\to 0}W\rho^{-1}\right|&<&\infty\;,\\[4mm]
\end{array}
\end{eqnarray}
where $M$ and $J$ are the gravitational mass and the angular momentum
of the star, respectively, see also Eqs (\ref{J_in}, \ref{M_in}) for an integral
representation. Therefore we write outside the star:
\begin{eqnarray}
  \nu &=& s\left[\nu_{\rm B}(t)+(s-1)H_\nu (s,t)\right]\\[3mm]
  \omega &=& s^3\left[\omega_{\rm B}(t)+(s-1) H_\omega (s,t)\right]\\[3mm]
   W^{\rm (ext)} &=& \rho\;\left(1+s^2\; \left[\hat{W}_{\rm
   B}(t)+(s-1)H_{W,\rm ext} (s,t)\right]\right)
\end{eqnarray}
and inside
\begin{eqnarray}
   U' &=& V_0+(s-1)H_{U'}(s,t)\\[3mm]
   a' &=& \rho^2\left[\hat{a}'_{\rm B}(t)+(s-1)H_{a'}(s,t)\right]\\[3mm]
   W^{\rm (int)} &=& \rho\;\left[1+\hat{W}_{\rm
   B}(t)+(s-1)H_{W,\rm int}(s,t)\right].
\end{eqnarray}
Here, the boundary values $a'_{\rm B},W_{\rm B}$ of the potentials $a',W$ 
are expressed by the functions $\hat{a}'_{\rm B},\hat{W}_{\rm B}$ in the
following manner:
\begin{eqnarray}
W_{\rm B}&=&\rho(\hat{W}_{\rm B}+1)\\[3mm]
a_{\rm B}&=&\rho^2\;\hat{a}'_{\rm B}.
\end{eqnarray}
The above one- and two-dimensional functions are expressed as limits of 
Chebyshev-expansions,  e.g.~\footnote{The Chebyshev-polynomials are defined by
$T_j(x)=\cos[j\arccos(x)], x\in [-1,1]$.}
\begin{eqnarray}
\nu_B(t)&=&\lim\limits_{m\to\infty} \nu_B^{(m)}(t)\;,\\[3mm]
\nu_{\rm B}^{(m)}(t)&=&\sum\limits_{k=1}^m \nu_{\rm B}^{(k;m)} T_{k-1}(2t-1)\;,\\[7mm]
H_\nu (s,t)&=&\lim\limits_{m\to\infty} H_\nu^{(m)}(s,t)\;,\\[3mm]
H_\nu^{(m)} (s,t)&=&\sum\limits_{j,k=1}^m H_\nu^{(jk;m)}
T_{j-1}(2s-1)T_{k-1}(2t-1).
\end{eqnarray}
Similarly, taking into account the representation of the boundary in
(\ref{Surf_B}), we write the boundary function 
$y_{\rm B}$ as follows\footnote{In order to get $m$ unknowns representing the
surface of the star in the $m^{\rm th}$-order approximation discussed
in section \ref{App_Scheme}, we take the radii $r_{\rm e},r_{\rm p}$ and $(m-2)$ Chebyshev-coefficients for the
function $g$.}
\begin{eqnarray}
y_{\rm B}&=&(1-t)\;[1+r_{\rm p}^{-2}tg(t)],\\[3mm]
g(t)&=&\lim\limits_{m\to\infty} g^{(m)}(t)\;,\\[3mm]
g^{(m)}(t)&=&\sum\limits_{k=1}^{m-2} g^{(k;m)} T_{k-1}(2t-1).
\end{eqnarray}
In the order $m$ of our approximation scheme, we establish a nonlinear 
set of algebraic equations that determines the above coefficients of
the $m^{\rm th}$ Chebyshev-expansion. In the limit $m\to\infty$, this 
set of algebraic equations is equivalent to the free boundary value problem in
question, and the $m^{\rm th}$ approximation becomes the solution.

\subsection{The nonlinear set of algebraic equations}
\label{App_Scheme}
For a given equation of state, we specify the solution of our free
boundary value problem by the prescription of a particular parameter pair. At
first let us take ($V_0,\Omega$); a more general choice will be discussed in
section \ref{Par_Prescr}.

We express the boundary values $\nu_{\rm B}$ and $\omega_{\rm B}$ in terms of 
$(V_0, \Omega)$ and the functions $a'_{\rm B}$ and $W_{\rm B}$, see Eqs
(\ref{Transf}). This ensures the continuity conditions of the field potentials at the
star's surface\footnote{The continuity conditions of the fields' derivatives
will be part of the set of algebraic equations in question.}.
Hence, in the order $m$ of our approximation scheme, we take 
the two-dimensional Chebyshev-coefficients
\begin{eqnarray}H^{(jk;m)}_\nu, H^{(jk;m)}_\omega, H^{(jk;m)}_{W,\rm ext}, H^{(jk;m)}_{U'}, 
H^{(jk;m)}_{a'}, H^{(jk;m)}_{W,\rm int}\end{eqnarray}
as well as the one-dimensional Chebyshev-coefficients 
\begin{eqnarray}(\hat{a}'_{\rm B})^{(k;m)}, \hat{W}_{\rm B}^{(k;m)}, g^{(k;m)}\end{eqnarray} 
as independent variables. They build up a vector ${\bf x}^{(m)}$ consisting of 
\begin{eqnarray}m_{\rm total}=6 m^2+3m\end{eqnarray}
components. The first $6m^2$ components comprise all two-dimensional 
Chebyshev-coefficients while the following $3m-2$ are the above
one-dimensional Chebyshev-coefficients. The remaining two entries are filled by
the values of $r_e$ and $r_p$. 

We now describe in detail the components of a vector 
\begin{eqnarray}{\bf f}^{(m)}={\bf f}^{(m)}({\bf x}^{(m)})\end{eqnarray} also consisting of
$m_{\rm total}$ components that must vanish for the solution ${\bf x}^{(m)}$ of 
the $m^{\rm th}$-order approximation.

Given an arbitrary vector ${\bf x}^{(m)}$, we compute the Chebyshev
coefficients corresponding to the first and second derivatives of 
the functions\footnote{\label{Cheb_Der} 
Note that it is straightforward to calculate (i) the Chebyshev
coefficients of a function from its values at the gridpoints $(s_j,t_k)$ (or
$t_k$ in the one-dimensional case), see Eq. (\ref{Zeros}), (ii) the value of a
function at an arbitrary point inside or at the boundary of $I^2$ (or $I$) 
from its Chebyshev coefficients, and (iii) the Chebyshev
coefficients of the derivative and the integral of a function from its Chebyshev
coefficients, see Press et al. (\cite{Num_Rec}).}
\begin{equation}\label{H_func}
H^{(m)}_\nu, H^{(m)}_\omega, H^{(m)}_{W,\rm ext}, H^{(m)}_{U'}, 
H^{(m)}_{a'}, H^{(m)}_{W,\rm int}\end{equation}
and
\begin{eqnarray}(\hat{a}'_{\rm B})^{(m)}, \hat{W}_{\rm B}^{(m)}, g^{(m)}\end{eqnarray} 
with respect to $s$ and $t$. Together with the coordinate transformations
(\ref{Transf_Int},\ref{Transf_Ext}), 
we therewith find the first and second spatial derivatives of all gravitational
potentials with respect to the coordinates $\rho$ and $\zeta$ in our 
$m^{\rm th}$-order approximation, at an arbitrary grid point inside the domains (not at
the origin or at infinity). So, we may fill the first $3m^2$ entries of 
${\bf f}^{(m)}$ with the differences of right and left hand sides of the exterior
equations (\ref{Exterior}), evaluated at $m^2$ gridpoints $(s_j,t_k), j,k=1\ldots m$,
corresponding to spatial points outside the star. Following the spirit of the
spectral methods and in order to ensure a rapid
convergence, we take for the $s_j$ and $t_j$ the roots of the 
$m^{\rm th}$ Chebyshev polynomial, i.~e.~
\begin{equation}\label{Zeros}
  s_j=t_j=\cos^2\left(\pi\,\frac{2j-1}{4m}\right)\quad(s_j,t_j>0).\end{equation}
For the subsequent $3m^2$ components of ${\bf f}^{(m)}$, we first compute the
two-dimensional, $m^{\rm th}$-order Chebyshev coefficients corresponding to the
interior function $k'$. This is done by determining the Chebyshev coefficients
corresponding to the $t$-derivative of $k'$ using (\ref{k_prime}) and again the coordinate
transformation (\ref{Transf_Int}), and after that, by integrating with respect to the
axis condition (\ref{k_axis}), see footnote \ref{Cheb_Der}. So, the $3m^2$ entries in
question can now be filled with the differences of right and left hand sides of
the interior equations (\ref{Interior}), again evaluated at the $m^2$ gridpoints $(s_j,t_k)$
here corresponding to spatial points inside the star.

The remaining $3m$ components of our vector ${\bf f}^{(m)}$ are formed from
the differences of the $m^{\rm th}$-order interior and exterior 
normal derivatives of the gravitational potentials $\nu,\omega$ and $W$,
evaluated at the $m$ surface grid points $(s=1,t_k)$. Note that the coordinate
transformations (\ref{Transf_Int},\ref{Transf_Ext}) are regular here, 
and thus the normal derivatives can easily be
computed using the shape of the star that is incorporated in ${\bf x}^{(m)}$. The
interior normal derivatives of $\nu$ and $\omega$ follow from the transformation 
formulae (\ref{Transf}) and the interior potentials.

In this manner we get in the $m^{\rm th}$ approximation order 
a nonlinear set of $m_{\rm total}=6 m^2+3m$ algebraic equations 
\begin{equation}\label{System}
{\bf f}^{(m)}({\bf x}^{(m)})=0\end{equation}
that is solved by a Newton-Raphson  method, see section \ref{Newton_Raphson}.

\subsection{The Newton-Raphson method and the initial solution}
\label{Newton_Raphson}

In the Newton-Raphson method, the zero of a
nonlinear set of algebraic equations of the form (\ref{System}) is determined
iteratively, 
\begin{equation}\label{New-Raph}
{\bf x}^{(m)}_n={\bf x}^{(m)}_{n-1}
   -\left[\left.\frac{\partial\, {\bf f}^{(m)}}{{\partial\, {\bf x}}^{(m)}}\right|_{{\bf
   x}^{(m)}_{n-1}}\right]^{\,-1} {\bf f}^{(m)}({\bf x}^{(m)}_{n-1})\,,\end{equation}
requiring an initial ${\bf x}^{(m)}_0$ which must already be sufficiently
close to the final solution ${\bf x}^{(m)}=\lim_{n\to\infty}{\bf x}^{(m)}_n$.
The Jacobi matrix in the Eq. (\ref{New-Raph}) is determined
approximately using ($\epsilon\ll 1$) 
\begin{eqnarray}\label{Jacobi}
\left[\left.\frac{\partial\, {\bf f}^{(m)}}{{\partial\, {\bf x}}^{(m)}}
      \right|_{{\bf x}^{(m)}}\right]_{AB}\approx
	\frac{1}{2\epsilon}
   \left[{\bf f}_A^{(m)}\left({\bf x}^{(m)}+\epsilon\,{\bf e}_B\right)-
         {\bf f}_A^{(m)}\left({\bf x}^{(m)}-\epsilon\,{\bf
	   e}_B\right)\right].
\end{eqnarray}
Here the subscripts $A$ and $B$ denote the corresponding element of the Jacobi
matrix and the vector ${\bf f}^{(m)}$, and ${\bf e}_B$
is the $B^{\rm th}$ unit vector, $({\bf e}_B)_A=\delta_{AB}$.

There are various possibilities for obtaining an initial solution. For example,
one may start from the static solution characterized by $\Omega=0$. Here the
corresponding field equations turn into ordinary differential equations with
respect to the radial coordinate $r$, and these can be solved e.g.~by using a
Runge-Kutta-method. Taking this solution for the initial ${\bf x}^{(m)}_0$, one may
now gradually increase the parameter $\Omega$ and thus eventually explore the
whole parameter space. Another possibility is to start with a Newtonian solution 
(e.g.~a Maclaurin spheroid). Proceeding into the 
relativistic regime comes about by increasing the absolute value of $V_0$.

In our treatment we favoured the latter initialization, for it already provides 
highly distorted bodies. Moreover, we calculated configurations
with a particular equation of state by starting from a constant mass-energy density
profile and continuously moving to the desired equation of state. 

\subsection{Arbitrary parameter prescription}
\label{Par_Prescr}
With a slight modification of our nonlinear set of equations described in section
\ref{App_Scheme}, we are able to take various different parameter prescriptions
into account. The idea is to add the quantities $\Omega$ and $V_0$ to the
vector ${\bf x}^{(m)}$, resulting in $m_{\rm total}=6m^2+3m+2$ unknowns from
now on. Simultaneously, we add two equations to the nonlinear set representing
exactly the desired parameter relation for the solution in question. This can
be done since all physical quantities concerning the final solution are now
contained in the vector ${\bf x}^{(m)}$. 

For example the potential $U'_{\rm c}$ at the origin reads\footnote{\label{H_const} 
Note that the function $H_{U'}^{(m)}(s=0,t)$ tends to a 
constant in the limit $m\to
\infty$. Similar properties hold for all functions listed in
(\ref{H_func}), see also section \ref{Uniqueness}.}
\begin{eqnarray}U'_{\rm c}=V_0-H_{U'}(0,0)\;,\end{eqnarray}
which is directly connected to the central pressure $p_{\rm c}$, 
see Eq. (\ref{Int_Eul}). 
Likewise, $M$ and $J$ can be expressed 
  \begin{eqnarray}
  M&=& - r_{\rm p}\left[\nu_{\rm B}(0)-H_\nu (0,0)\right]\;,\\[3mm]
  J&=& \frac{1}{2}r_{\rm p}^3\left[\omega_{\rm B}(0)- H_\omega
  (0,0)\right].
  \end{eqnarray}
Also the prescription of a parameter $\beta$ is possible which controls the
distance of a configuration to the mass shedding limit:
\begin{equation}\label{beta}\beta=-\frac{d y_{\rm B}}{dt}\;(t=1)=\left\{
          \begin{array} {cl} 0  & \mbox{in the mass shedding limit} \\
                             1 & \mbox{for an ellipsoidal shape}
				     \end{array}\right.
 \end{equation}
Similarly, one can prescribe more complicated expressions such as the
baryonic mass $M_0$, which is defined by an integral 
over the interior field quantities.

Any two conditions of this kind (of which the above ones are just examples) can be
taken and added to the system of nonlinear equations. The corresponding
parameters (here $p_{\rm c}, M,J,\beta$ or $M_0$) must then be prescribed. 
In this paper we concentrate on the pair 
$(p_{\rm c},r_{\rm p}/r_{\rm e})$ and only take $(p_{\rm c},\beta)$
in order to place ourselve exactly on the mass shedding limit, see section \ref{Shed}.

\subsection{Regularity and uniqueness}
\label{Uniqueness}

As already depicted in section \ref{Mappings}, the AKM-method is characterized
by the fact that some of the usual boundary conditions are replaced by
regularity requirements. Moreover, if for the moment we only consider the functions 
\begin{equation}\label{H_func1}
H_\nu, H_\omega, H_{W,\rm ext}, H_{U'}, 
H_{a'}, H_{W,\rm int}\end{equation}
and treat the quantities $r_{\rm e},r_{\rm p},\Omega,V_0$ as well as the 
functions $a'_{\rm B}, W_{\rm B}$ and $g$ 
as if they were given, we obtain specific partial differential
equations valid in $I^2$, and particular boundary conditions at any edge of
$I^2$ are not required for any of the functions listed in
(\ref{H_func1}). Nevertheless, the solution of this system of equations is
uniquely determined if we require regularity with respect to all functions. 
A similar situation can be studied in the one-dimensional case,
e.~g.~the equation ($\;\dot{}= d/dt$)
\begin{eqnarray}
t(1-t)\ddot{h}+2(1-2t)\dot{h}-2h+2=0\Leftrightarrow
 [t(1-t)h]^{\cdot\cdot}=-2\end{eqnarray}
possesses only the solution $h\equiv 1$ which is regular within $I$. 

The above approximation scheme sorts out the non-regular solutions
since it is based on Chebyshev-expansions. It moreover ensures known, additional
properties of the functions (\ref{H_func1}) at $s=0$,
e.~g.~$H_{U'}(s=0,t)=\rm const$. Note that these properties are approached as
$m\to \infty$. 

\section{Representative Examples}
\label{Examples}

\subsection{Ordinary stars}
\label{Ordinary}

\begin{table*}
\begin{center}
   \begin{tabular}{lllllllllll}\hline
     \multicolumn{2}{c}{}&\multicolumn{1}{c}{m=6}&
     \multicolumn{1}{c}{m=8}&\multicolumn{1}{c}{m=10}
     &\multicolumn{1}{c}{m=12}&\multicolumn{1}{c}{m=14}
     &\multicolumn{1}{c}{m=16}&\multicolumn{1}{c}{m=18}
     &\multicolumn{1}{c}{m=20}&\multicolumn{1}{c}{m=22} \\ \hline
$\bar{p}_{\rm c}$     &    1            &        &        &        &        & 
                                                 &        &        &        &  \\
$r_{\rm p}/r_{\rm e}$ & $0.7          $ &        &        &        &        & 
                                                 &        &        &        &  \\  
$\bar{\Omega}$        & $1.41170848318$  & 1.9e-04 & 1.3e-05 & 7.8e-07 & 2.9e-08 & 
                                           8.5e-10 & 4.6e-11 & 3.0e-12 & 1.3e-13 &
							 8.0e-15 \\
$\bar{M}$             & $0.135798178809$ & 1.8e-04 & 3.5e-06 & 5.9e-08 & 3.4e-09 & 
                                           3.8e-10 & 2.6e-11 & 8.5e-13 & 3.3e-14 &
							 6.8e-15 \\
$\bar{R}_{\rm circ}$&$0.345476187602$   &  2.0e-08 & 1.5e-06 & 1.7e-08 & 1.8e-09 & 
                                          4.2e-11 & 1.8e-11 & 1.6e-12 & 1.1e-13 &
							1.3e-14 \\
$\bar{J}$&$0.0140585992949$             & 8.7e-04 & 6.8e-05 & 3.7e-06 & 1.2e-08 & 
                                          1.2e-08 & 6.8e-10 & 8.4e-12 & 3.5e-12&
							2.0e-13 \\
$Z_{\rm p}$&$1.70735395213$             & 3.2e-05 & 6.5e-06 & 2.4e-07 & 3.6e-09 & 
                                          4.6e-10 & 9.1e-12 & 7.1e-13 & 1.7e-13 &
							1.6e-14 \\[1mm] \hline 
$GRV2$&                                 & 7.5e-05 & 3.9e-06 & 3.9e-07 & 2.2e-08 & 
                                          8.9e-10 &4.2e-11 & 3.1e-12 & 3.0e-13 &
							7.7e-14 \\ 
$GRV3$&                                 & 1.2e-05 & 7.5e-06 & 1.2e-07 & 2.9e-08 & 
                                          1.4e-09 &3.5e-11 & 1.3e-12 & 1.8e-13 &
							6.5e-14 \\
$|1-M_{\rm in}/M_{\rm out}|$&           &  2.8e-04 & 4.9e-06 & 1.9e-07 & 1.1e-08 & 
                                           4.1e-10 &3.4e-12 & 1.5e-12 & 4.2e-13 &
							 2.3e-13 \\
$|1-J_{\rm in}/J_{\rm out}|$&           &  1.2e-03 & 7.0e-05 & 4.1e-06 & 5.0e-08 & 
                                           1.1e-08 &7.1e-10 & 4.6e-12 & 2.9e-12 &
							 1.1e-13 \\
          \hline\vspace*{2mm}
         \end{tabular}
         \caption{Results for a constant mass-energy density
           model ($\mu=\mu_0$) with $\bar{p}_{\rm c}=1$, $r_{\rm p}/r_{\rm e}=0.7$. 
	     Here, $\bar{p}_{\rm c}=p_{\rm c}/\mu_0$, 
           $\bar{\Omega}=\Omega/\mu_0^{1/2}$, $\bar{M}=M \mu_0^{1/2}$,
           $\bar{R}_{\rm circ}={R}_{\rm circ}\,\mu_0^{1/2}$ and $\bar{J}=J\mu_0$ are
           normalized values of the physical quantities, 
	     see Eqs (\ref{J_in}, \ref{M_in}). Apart from
	     the virial identities $GRV2$ and $GRV3$ in the $m^{\rm th}$ order
	     approximation, the columns 3-11 display the relative deviation of the
	     specific quantity in the $m^{\rm th}$ order
	     approximation with respect to the numerical result obtained for
	     $m=24$.
	     The quantities $M_{\rm in},J_{\rm in}$ and $M_{\rm
	     out},J_{\rm out}$ refer to the corresponding numerical values
	     resulting from (\ref{J_in}, \ref{M_in}) and (\ref{MJ_out}) respectively.}
   \end{center}
\end{table*}
\begin{table*}
\begin{center}
   \begin{tabular}{lllllllllll}\hline
     \multicolumn{2}{c}{}&\multicolumn{1}{c}{m=6}&
     \multicolumn{1}{c}{m=8}&\multicolumn{1}{c}{m=10}
     &\multicolumn{1}{c}{m=12}&\multicolumn{1}{c}{m=14}
     &\multicolumn{1}{c}{m=16}&\multicolumn{1}{c}{m=18}
     &\multicolumn{1}{c}{m=20}&\multicolumn{1}{c}{m=22} \\ \hline
$\bar{\mu}_{\rm c}$     &    1         &        &        &        &        & 
                                                 &        &        &        &  \\
$r_{\rm p}/r_{\rm e}$ & $0.834        $ &        &        &        &        & 
                                                 &        &        &        &  \\  
$\bar{\Omega}$        & $0.4004385709$  & 1.1e-03 & 9.0e-05 & 7.3e-06 & 6.4e-07 & 
                                           6.4e-08 & 7.2e-09 & 8.6e-10 & 1.1e-10 &
							 1.3e-11 \\
$\bar{M}$             & $0.1605611357$ &  4.2e-04 & 5.7e-06 & 2.5e-06 & 3.2e-07 & 
                                           3.7e-08 & 4.1e-09 & 4.7e-10 & 5.4e-11 &
							 5.9e-12 \\
$\bar{R}_{\rm circ}$&   $0.6794279802$ &  5.7e-04 & 6.0e-05 & 5.4e-06 & 5.1e-07 & 
                                           5.2e-08 & 5.9e-09 & 7.0e-10 & 8.6e-11 &
							 9.8e-12 \\
$\bar{J}$&$0.009491087857$             &  9.4e-04 & 3.3e-05 & 8.5e-06 & 1.1e-06 & 
                                           1.2e-07 & 1.3e-08 & 1.4e-09 & 1.4e-10&
							 1.2e-11 \\
$Z_{\rm p}$&$0.4580590747$             &  1.7e-03 & 8.7e-05 & 4.6e-06 & 2.6e-07 & 
                                           1.9e-08 & 1.8e-09 & 2.0e-10 & 2.6e-11 &
							 3.1e-12 \\[1mm] \hline 
$GRV2$&                                 &  2.6e-04 & 8.1e-06 & 4.8e-07 & 7.5e-08 & 
                                           1.3e-08 &1.8e-09 & 2.4e-10 & 3.2e-11 &
							 4.3e-12 \\ 
$GRV3$&                                 &  5.5e-05 & 2.9e-06 & 2.7e-07 & 1.5e-08 & 
                                           1.2e-09 &2.0e-10 & 3.9e-11 & 7.1e-12 &
							 1.3e-12 \\
$|1-M_{\rm in}/M_{\rm out}|$&           &  1.2e-04 & 6.5e-06 & 5.5e-07 & 5.5e-08 & 
                                           5.5e-09 &5.3e-10 & 4.5e-11 & 2.4e-12 &
							 3.3e-13 \\
$|1-J_{\rm in}/J_{\rm out}|$&           &  2.7e-04 & 3.7e-05 & 4.0e-06 & 4.2e-07 & 
                                           4.1e-08 &3.6e-09 & 2.5e-10 & 5.6e-13 &
							 4.9e-12 \\
          \hline\vspace*{2mm}
         \end{tabular}
         \caption{Results for a polytropic
           model (polytropic exponent $\Gamma=2$, polytropic constant $K$) with 
	      $\bar{\mu}_{\rm c}=1$, $r_{\rm p}/r_{\rm e}=0.834$. 
	     Here, $\bar{\mu}_{\rm c}=K\mu_{\rm c}$, 
           $\bar{\Omega}=K^{1/2}\Omega$, $\bar{M}=K^{-1/2}M $,
           $\bar{R}_{\rm circ}=K^{-1/2}{R}_{\rm circ}$ and $\bar{J}=K^{-1}J$ are
           normalized values of the physical quantities, see Eqs
	     (\ref{J_in}, \ref{M_in}). For the meaning of the quantities listed in the
	     columns 3-11, see Table 1.}
\end{center}
\end{table*}
\begin{table*}
\begin{center}
   \begin{tabular}{lllllllllll}\hline
     \multicolumn{2}{c}{}&\multicolumn{1}{c}{m=6}&
     \multicolumn{1}{c}{m=8}&\multicolumn{1}{c}{m=10}
     &\multicolumn{1}{c}{m=12}&\multicolumn{1}{c}{m=14}
     &\multicolumn{1}{c}{m=16}&\multicolumn{1}{c}{m=18}
     &\multicolumn{1}{c}{m=20}&\multicolumn{1}{c}{m=22} \\ \hline
$\bar{p}_{\rm c}$     &    2            &        &        &        &        & 
                                                 &        &        &        &  \\
$r_{\rm p}/r_{\rm e}$ & $0.5          $ &        &        &        &        & 
                                                 &        &        &        &  \\  
$\bar{\Omega}$        & $3.4304996$   &  2.1e-04 & 1.5e-05 & 4.0e-07 & 1.6e-07 & 
                                           6.2e-08 & 1.9e-08 & 5.3e-09 & 1.5e-09 &
							 3.4e-10 \\
$\bar{M}$             & $0.035510326$ &  5.3e-03 & 5.9e-04 & 6.4e-05 & 7.2e-06 & 
                                           9.2e-07 & 1.4e-07 & 2.6e-08 & 5.8e-09 &
							 1.3e-09 \\
$\bar{R}_{\rm circ}$&   $0.14117783$ &   3.3e-05 & 3.2e-09 & 1.8e-07 & 8.4e-08 & 
                                           3.0e-08 & 9.4e-09 & 2.9e-09 & 8.2e-10 &
							 2.0e-10 \\
$\bar{J}$&$0.0011024838$              &  1.6e-03 & 2.8e-04 & 5.4e-05 & 8.5e-06 & 
                                           1.4e-06 & 2.8e-07 & 6.4e-08 & 1.6e-08&
							 3.5e-09 \\
$Z_{\rm p}$&$0.72634557$              &  8.6e-05 & 1.8e-05 & 5.2e-06 & 1.3e-06 & 
                                           3.2e-07 & 8.4e-08 & 2.4e-08 & 6.3e-09 &
							 1.5e-09 \\[1mm] \hline 
$GRV2$&                                 &  1.2e-04 & 4.8e-06 & 4.1e-07 & 4.0e-08 & 
                                           3.9e-09 &3.9e-10 & 3.8e-11 & 3.2e-12 &
							 4.7e-14 \\ 
$GRV3$&                                 &  6.7e-04 & 3.9e-05 & 2.8e-06 & 2.4e-07 & 
                                           2.1e-08 &1.8e-09 & 1.5e-10 & 1.1e-11 &
							 5.2e-14 \\
$|1-M_{\rm in}/M_{\rm out}|$&           &  5.9e-03 & 6.0e-04 & 6.2e-05 & 6.5e-06 & 
                                           6.9e-07 &7.4e-08 & 7.9e-09 & 8.4e-10 &
							 8.9e-11 \\
$|1-J_{\rm in}/J_{\rm out}|$&           &  2.2e-03 & 2.7e-04 & 4.6e-05 & 6.0e-06 & 
                                           6.5e-07 &8.1e-08 & 1.1e-08 & 1.2e-09 &
							 1.5e-10 \\
          \hline\vspace*{2mm}
         \end{tabular}
         \caption{Results for a strange star 
           model (MIT bag constant $B$) with 
	     $\bar{p}_{\rm c}=2$, $r_{\rm p}/r_{\rm e}=0.5$. 
	     Here, $\bar{p}_{\rm c}=B^{\,-1}p_{\rm c}$, 
           $\bar{\Omega}=B^{\,-1/2}\Omega$, $\bar{M}=B^{1/2}M $,
           $\bar{R}_{\rm circ}=B^{1/2}{R}_{\rm circ}$ and $\bar{J}=BJ$ are
           normalized values of the physical quantities, see Eqs
	     (\ref{J_in}, \ref{M_in}). For the meaning of the quantities listed in the
	     columns 3-11, see Table 1.}
\end{center}
\end{table*}

At first we apply the AKM-method to three models of homogeneous 
[Eqs (\ref{EOS_hom})], polytropic
[with a polytropic exponent $\Gamma=2$, Eqs (\ref{EOS_pol})], and strange stars [Eqs (\ref{EOS_str})]. 
In particular, we prescribe the corresponding equation of state in the form
$\mu=\mu(p)$ and find from Eq. (\ref{Int_Eul}) the relation to the
interior potential $U'$:
\begin{eqnarray}
\label{EOS_hom}
\mu\,(p)=\mu_0={\rm const}&\Rightarrow&\left\{
               \begin{array}{lcl}
		       p&=&\mu_0\left({\rm e}^{V_0-U'}-1\right) \\[3mm]
		       \mu&=&\mu_0
               \end{array}\right.\\[5mm]
\label{EOS_pol}
\mu\,(p)=p+\sqrt{p/K}&\Rightarrow&\left\{
               \begin{array}{lcl}
		       p&=&\left({\rm e}^{V_0-U'}-1\right)^2/(4K) \\[3mm]
		       \mu&=&\left({\rm e}^{2\,(V_0-U')}-1\right)/(4K)
		   \end{array}\right.\\[5mm]
\label{EOS_str}
\mu\,(p)=4B+3p&\Rightarrow&\left\{
               \begin{array}{lcl}
		       p&=&B\left({\rm e}^{4\,(V_0-U')}-1\right) \\[3mm]
		       \mu&=&B\left(1+3{\rm e}^{4\,(V_0-U')}\right)
		   \end{array}\right.
\end{eqnarray}
Here, $K$ and $B$ are the polytropic constant and the MIT bag
constant, respectively\footnote{For a description of the equations of
state corresponding to polytropic and strange star matter, see Tooper
(\cite{Too65}) and e.g.~Gourgoulhon et al. (\cite{gou99}) respectively.}.
Note that for the application of the AKM-method (strictly speaking, only for its
rapid convergence) it is necessary to have analytic dependencies 
$p=p(U')$ and $\mu=\mu(U')$, in
particular at $U'=V_0$. For the equation of state, $\mu=\mu(p)$, this requires that 
\begin{eqnarray}\mu=p^{N/(N+1)}f[p^{1/(N+1)}]\end{eqnarray} 
where $f$ is some function which is
positive and analytic when its argument vanishes, and $N$ is some non-negative integer.
Apart from the homogeneous and the strange star model, this condition is
met for polytropic equations of state with a polytropic exponent
$\Gamma=1+1/N$ when $N$ is a non-negative integer (as in the case above where
$N=1$). In order to treat more general equations of state,
one needs to consider several layers inside the star, with each one of them 
characterized by a specific equation of state. The outermost one of them again 
must meet the above requirement. The consideration of several layers leads to
a corresponding number of subregions into which the interior domain needs to be
split. 

In Tables 1 to 3 one finds numerical values of several physical quantities,
for a specified configuration with prescribed central pressure $p_{\rm c}$
(equivalently, for non-homogeneous models, we may prescribe the central
mass-energy density $\mu_{\rm c}$, see Table 2) and
radius ratio $r_{\rm p}/r_{\rm e}$. The angular
momentum $J$, gravitational mass $M$, equatorial circumferential radius 
${R}_{\rm circ}$ and the polar redshift $Z_{\rm p}$ 
are given by :
\begin{eqnarray} 
\label{J_in}
J&=&-2\pi\int(\mu+p)a'{\rm e}^{2k'-2U'}Wd\rho d\zeta\\[4mm]
\label{M_in}
M&=&2\Omega J+2\pi\int(\mu+3p){\rm e}^{2k'-2U'}Wd\rho d\zeta
\end{eqnarray}
\begin{eqnarray} 
\begin{array}{lcl}
  R_{\rm circ}&=&{\rm e}^{-V_0}\left[W\sqrt{1-u^2}\,\right]_{\,(\rho=r_{\rm
  e},\,\zeta=0)}\\[4mm]
  Z_{\rm p}&=&{\rm e}^{-V_0}-1
\end{array}
\end{eqnarray}
with $u=-W^{-1}a'{\rm e}^{2U'}$. Note that the above integrals extend over
the space of $\rho$- and $\zeta$-coordinates covering the interior of the 
body\footnote{The quantities $M$ and
$J$ can also be taken from the exterior fields $\nu$ and $\omega$, see Eqs
\ref{MJ_out}.}.

A first test of the accuracy of a solution determined numerically
is the comparison of the calculations of $M$ and
$J$ from the exterior fields [see Eqs (\ref{MJ_out})]
with those from the above integral representations
(\ref{J_in}, \ref{M_in}). A further check is given by  
the general-relativistic virial identies $GRV2$ and $GRV3$, derived by Bonazzola \&
Gourgoulhon \cite{bonaz} and Gourgoulhon \& Bonazzola \cite{gourg}. As a
consequence of the field equations, they identically vanish for an exact
analytic solution corresponding to a stationary and
asymptotically flat spacetime. Particularly, for our rotating star models they
read 
\begin{eqnarray} GRV3=|1-I_1/I_2|\;,\quad GRV2=|1-J_1/J_2|\end{eqnarray}
with :
\begin{eqnarray} \label{GRV}
  I_1&=&4\pi\int\left[3p\sqrt{1-u^2}+(\mu+p)\frac{u^2}{\sqrt{1-u^2}}\right]
       {\rm e}^{2k'-3U'}Wd\rho d\zeta \\[3mm]
 I_2&=&\int{\rm e}^\eta\left\{\rho\left[\left(\nabla\nu\right)^2-\frac{1}{2}\nabla\alpha\nabla\eta
       -\frac{3}{8}W^2{\rm e}^{-4\nu}\left(\nabla\omega\right)^2\right]\right.
	  \left.-\frac{1}{2}\left(1-{\rm e}^{2(\alpha-\eta)}\right)
     \left(\alpha_\rho-\frac{1}{2}\,\eta_\rho\right)\right\}
     d\rho d\zeta\\
 J_1&=&8\pi\int\left[p+(\mu+p)\frac{u^2}{1-u^2}\right]
       {\rm e}^{2k'-2U'}d\rho d\zeta \\[3mm]  
 J_2&=&\int\left[\left(\nabla\nu\right)^2
        -\frac{3}{4}W^2{\rm e}^{-4\nu}\left(\nabla\omega\right)^2\right]d\rho
	  d\zeta 	  
\end{eqnarray}
(we use the abbreviations $u=-W^{-1}a'{\rm e}^{2U'},\,\rho{\rm e}^{\eta}=W{\rm
e}^{-\nu}$). 
The Nabla-operator has its usual meaning, in terms of the coordinates
$\rho$ and $\zeta$, i.~e.~for any two functions $a$ and $b$ 
\begin{eqnarray}\left(\nabla a\right)^2=a_\rho^2+a_\zeta^2\;,\quad 
  \nabla a\nabla b=a_\rho b_\rho+a_\zeta b_\zeta.\end{eqnarray}
The above integrals are taken over the whole space of $\rho$- and $\zeta$-coordinates 
(for $I_1$ and $J_1$ this reduces again to the interior region of the star 
since both pressure $p$ and energy-mass density $\mu$ vanish
outside the body). 

Apart from the values for the above physical quantities with the accuracy that was
reached in the $24^{\rm th}$ approximation order, we provide in Tables 1-3 the 
improvement of the accuracy as the order $m$ is increased. 
Also given are the corresponding numerical values of the general-relativistic virial identies 
$GRV2$ and $GRV3$ as well as those of the relative deviations concerning the integral
and far-field representations of $M$ and $J$. 

We note generally an exponential convergence of the numerical solution as the
order $m$ increases. This is a common feature of the spectral methods. However, the
star's field quantities may vary in their `smoothness' resulting in a variably
rapid convergence. For example, the convergence of the numerical solution
corresponding to the homogeneous star (Table
1) is much faster than that corresponding to the strange one (Table 3). 
In a sense, the strange star is closer to the mass-shedding limit (here
$\beta\approx 0.39$ while for the homogeneous model $\beta\approx0.84$) and
moreover more flattened. 

The models in Tables 1 and 2 have been calculated by Nozawa et al. (\cite{nozawa}).
Note that for the polytropic model there is a steeper gradient 
of the pressure as a function of the radial coordinate $r$, e.g.~within the equatorial
plane. In order to take this property into account, we used for this model the
following slightly modified interior coordinate transformation
\begin{eqnarray} \label{Transf_Int1}
\rho^2=r_{\rm e}^2\hspace*{0.1mm}\sigma(s\,;c_{\rm s})\hspace*{0.1mm}t\,,\quad 
\zeta^2=r_{\rm p}^2\hspace*{0.1mm}\sigma(s\,;c_{\rm s})\hspace*{0.1mm}y_{\rm B}(t)\,,\quad (s,t)\in I^2
\end{eqnarray}
with \begin{eqnarray}\sigma(s\,;c_{\rm s})=\frac{1-c_{\rm s}}{1-c_{\rm s}s}s\end{eqnarray} 
and the constant parameter $c_{\rm s}=0.6$ (for the
other models we took $c_{\rm s}=0$). Minor modifications of this kind, 
specially suited to the particular problem in question,
may accelerate the convergence, see below for further examples.

\subsection{Critical stars}
\label{Critical}

\subsubsection{Stars at the mass shedding limit}
\label{Shed}

\begin{figure}
\unitlength1cm
\hspace*{-1cm}
\begin{center}
\psfrag{r}[r][r]{{$\rho$}}
\psfrag{q}[r][r]{{$\zeta$}}
\epsfig{file=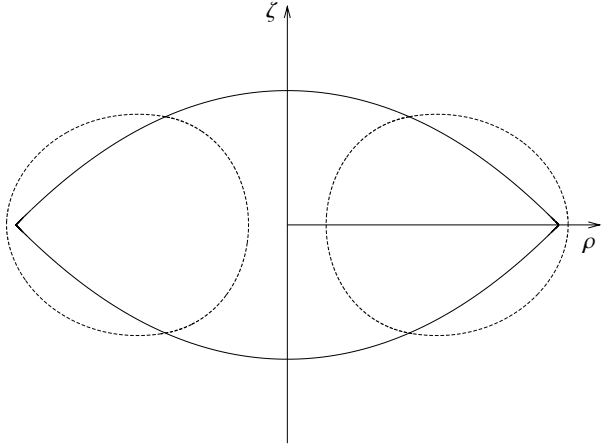,scale=0.7}
\end{center}
\caption{Meridional cross-section of the homogeneous mass-shedding
configuration specified in Table 4 (with the axes scaled identically). 
The dashed curves indicate the boundary of
the corresponding ergo-region.}
\end{figure}
\begin{figure}
\unitlength1cm
\hspace*{-1cm}
\begin{center}
\psfrag{t}[r][r]{{$t$}}
\psfrag{gB}[c][c]{{ $dg/dt$}}
\psfrag{0}[r][r]{{ $0$}}
\psfrag{1}[c][c]{{ $1$}}
\psfrag{-0.02}[c][c]{{}}
\psfrag{-0.04}[c][c]{{}}
\psfrag{-0.06}[c][c]{{}}
\psfrag{-0.08}[c][c]{{}}
\psfrag{-0.1}[r][r]{{ $-0.1$}}
\psfrag{-0.12}[c][c]{{}}
\psfrag{-0.14}[c][c]{{}}
\psfrag{-0.16}[c][c]{{}}
\psfrag{0.2}[c][c]{{0.2}}
\psfrag{0.4}[c][c]{{0.4}}
\psfrag{0.6}[c][c]{{0.6}}
\psfrag{0.8}[c][c]{{0.8}}
\epsfig{file=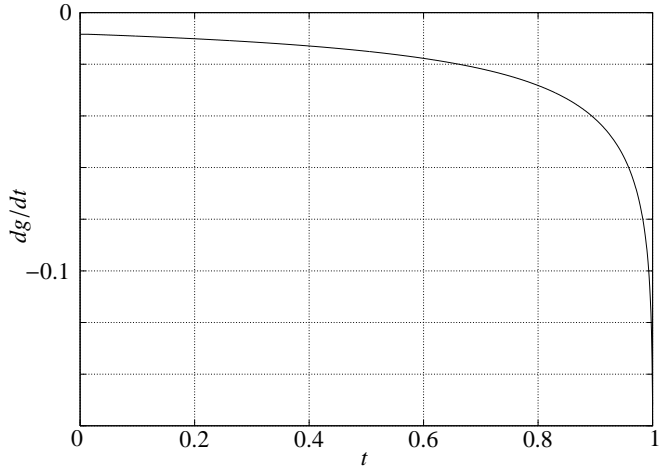,scale=0.7}
\end{center}
\caption{First derivative of the function $g=g(t)$ with respect
to $t$ for the homogeneous Newtonian mass-shedding configuration (A) of table 2
in Ansorg, Kleinw\"{a}chter \& Meinel (\cite{AKM03b}). The numerical methods
explained in section 3 ibid. have been carried out up to the approximation
order $N=80$.}
\end{figure}

\begin{table}
\begin{center}
   \begin{tabular}{lllll}\hline
     \multicolumn{2}{c}{}&\multicolumn{1}{c}{m=10}&
     \multicolumn{1}{c}{m=16}&\multicolumn{1}{c}{m=22}\\ \hline
$\bar{p}_{\rm c}$     &   1            &          &         &         \\
$r_{\rm p}/r_{\rm e}$ & $0.4918$       &  7.2e-04 & 1.2e-04 & 1.6e-05 \\
$\bar{\Omega}$        & $1.6588$       &  6.1e-05 & 5.5e-06 & 5.8e-07 \\
$\bar{M}$             & $0.1623$       &  7.2e-05 & 3.8e-06 & 6.1e-07 \\
$\bar{R}_{\rm circ}$&   $0.4041$       &  3.2e-05 & 7.1e-06 & 1.2e-06 \\
$\bar{J}$             & $0.02431$      &  1.8e-04 & 9.4e-06 & 1.6e-06 \\
$Z_{\rm p}$           & $2.6685$       &  1.2e-04 & 1.6e-05 & 2.1e-06 \\[1mm] \hline 
$GRV2$&                                &  1.2e-04 & 1.6e-05 & 4.3e-06 \\ 
$GRV3$&                                &  1.3e-04 & 1.8e-05 & 4.9e-06 \\
$|1-M_{\rm in}/M_{\rm out}|$&          &  2.8e-04 & 4.9e-05 & 1.4e-05 \\
$|1-J_{\rm in}/J_{\rm out}|$&          &  4.2e-04 & 8.1e-05 & 2.3e-05 \\
          \hline\vspace*{2mm}
         \end{tabular}
         \caption{Results for a constant mass-energy density
           model rotating at the mass-shedding limit 
	     with $\bar{p}_{\rm c}=1$. For the meaning of the quantities, see
	     Table 1.}
\end{center}
\end{table}
\begin{table}
\begin{center}
   \begin{tabular}{lllll}\hline
     \multicolumn{2}{c}{}&\multicolumn{1}{c}{m=10}&
     \multicolumn{1}{c}{m=16}&\multicolumn{1}{c}{m=22}\\ \hline
$\bar{\mu}_{\rm c}$   &   0.34         &          &         &         \\
$r_{\rm p}/r_{\rm e}$ &   0.5845178    &  9.1e-05 & 9.7e-07 & 3.6e-08 \\
$\bar{\Omega}$        &   0.3770150    &  2.0e-04 & 7.0e-07 & 3.8e-09 \\
$\bar{M}$             &   0.1883522    &  8.7e-05 & 9.7e-08 & 1.1e-09 \\
$\bar{R}_{\rm circ}$  &   1.0920220    &  3.4e-05 & 3.0e-07 & 2.6e-08 \\
$\bar{J}$             &   0.02023980   &  3.0e-05 & 1.4e-06 & 9.5e-09 \\
$Z_{\rm p}$           &   0.4035809    &  2.9e-04 & 6.0e-07 & 2.3e-09 \\[1mm] \hline 
$GRV2$&                                &  1.5e-05 & 8.7e-08 & 1.6e-09 \\ 
$GRV3$&                                &  5.5e-06 & 8.4e-08 & 1.3e-09 \\
$|1-M_{\rm in}/M_{\rm out}|$&          &  6.7e-05 & 2.7e-07 & 3.2e-09 \\
$|1-J_{\rm in}/J_{\rm out}|$&          &  2.8e-04 & 1.6e-06 & 1.8e-08 \\
          \hline\vspace*{2mm}
         \end{tabular}
         \caption{Results for a polytropic
           model (polytropic exponent $\Gamma=2$) at the mass-shedding limit with
	     $\bar{\mu}_{\rm c}=0.34$. For the meaning of the quantities, see Table 2.}
\end{center}
\end{table}

\begin{table}
\begin{center}
   \begin{tabular}{lllll}\hline
     \multicolumn{2}{c}{}&\multicolumn{1}{c}{m=10}&
     \multicolumn{1}{c}{m=16}&\multicolumn{1}{c}{m=22}\\ \hline
$\bar{p}_{\rm c}$     &   3          &          &         &         \\
$r_{\rm p}/r_{\rm e}$ &   0.4713     &  6.0e-04 & 1.4e-04 & 1.9e-05 \\
$\bar{\Omega}$        &   3.6505     &  4.0e-05 & 6.3e-06 & 6.0e-07 \\
$\bar{M}$             &   0.03719    &  1.5e-04 & 9.4e-06 & 1.3e-06 \\
$\bar{R}_{\rm circ}$  &   0.1444     &  3.0e-04 & 6.8e-05 & 9.8e-06 \\
$\bar{J}$             &   0.001205   &  1.3e-04 & 2.6e-05 & 3.7e-06 \\
$Z_{\rm p}$           &   0.82865    &  5.2e-05 & 1.6e-05 & 1.9e-06 \\[1mm] \hline 
$GRV2$&                              &  1.2e-04 & 1.8e-05 & 5.0e-06 \\ 
$GRV3$&                              &  1.6e-04 & 2.4e-05 & 6.5e-06 \\
$|1-M_{\rm in}/M_{\rm out}|$&        &  9.1e-05 & 3.5e-05 & 9.6e-06 \\
$|1-J_{\rm in}/J_{\rm out}|$&        &  4.6e-04 & 8.4e-05 & 2.4e-05 \\
          \hline\vspace*{2mm}
         \end{tabular}
         \caption{Results for a strange star 
           model at the mass-shedding limit with
	     $\bar{p}_{\rm c}=3$. For the meaning of the quantities, see Table 3.}
\end{center}
\end{table}

The endpoint of a sequence of rotating stars is often marked by a mass shedding 
limit. It is of particular interest since specific physical quantities such as the
angular velocity reach maximal values there. A highly accurate determination of this
limit is therefore desirable.

The mass shedding limit is reached when the angular velocity $\Omega$ of the star
attains the angular velocity of test particles moving on a prograde circular
orbit at the star's equatorial rim. For the $\rho$-derivative of the field 
quantity $U'$ it follows:
\begin{eqnarray}U'_{\rho}(r_{\rm e},0)=0.\end{eqnarray}
Moreover,  a cusp at the surface occurs (see Fig. 1), which
corresponds to a vanishing mass-shedding parameter $\beta$, see Eq. (\ref{beta}).

Numerical investigations of a homogeneous Newtonian configuration rotating at the 
mass-shedding limit suggest that the surface function $g$
becomes singular in higher derivatives at this limit, see Fig. 2.
This causes a similar singular
behaviour of all gravitational potentials, and we expect a failure of the
spectral methods. Nevertheless, since the singularities show up only in higher
derivatives, it is possible to achieve a slow convergence, see tables
4-6\footnote{Note that the example listed in Table 5 has previously been
calculated, see Nozawa et al. (\cite{nozawa}).}.
However, it is then necessary to modify the coordinate transformations
(\ref{Transf_Int},\ref{Transf_Ext}) such that the
curves $s={\rm const}$ do not possess a cusp (except for $s=1$). Here we
use 
\begin{eqnarray} \label{Transf_Int_Shed}
\rho^2=r_{\rm e}^2\hspace*{0.1mm}s\,t\,,\quad 
\zeta^2=s(1-t)\,[r_{\rm p}^2+stg(t)]\,,\quad (s,t)\in I^2
\end{eqnarray}
and
\begin{eqnarray} \label{Transf_Ext_Shed}
\rho^2=\frac{r_{\rm e}^2 t}{s^2}\,,\quad 
\zeta^2=\frac{(1-t)\,[r_{\rm p}^2+stg(t)]}{s^2}\,,\quad (s,t)\in I^2
\end{eqnarray}
for the interior and exterior region respectively.

From the numerical results listed in Tables 4-6 
we may speculate that the behaviour of the pressure at the star's surface (which
is determined by the equation of state) affects the type of the above
singularities. They seem to be weaker for smoother equations of state, when the
pressure and some higher derivatives vanish at the equator. 

\subsubsection{Stars with infinite central pressure}
\label{Pressure}
\begin{figure}
\unitlength1cm
\hspace*{-1cm}
\begin{center}
\psfrag{r}[r][r]{{$\rho$}}
\psfrag{q}[r][r]{{$\zeta$}}
\epsfig{file=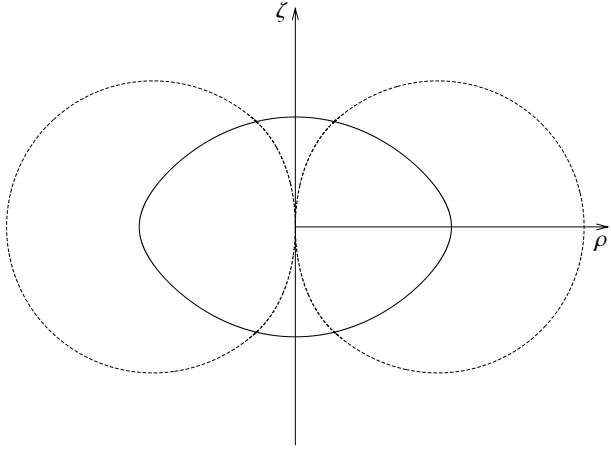,scale=0.7}
\end{center}
\caption{Meridional cross-section of a homogeneous configuration 
with infinite central pressure, specified in Table 7 
(with the axes scaled identically). 
The dashed curves indicate the boundary of
the corresponding ergo-region.}
\end{figure}

\begin{table}
\begin{center}
   \begin{tabular}{lllll}\hline
     \multicolumn{2}{c}{}&\multicolumn{1}{c}{m=10}&
     \multicolumn{1}{c}{m=16}&\multicolumn{1}{c}{m=22}\\ \hline
$\bar{p}_{\rm c}$     &   $\infty$   &   &  & \\
$r_{\rm p}/r_{\rm e}$ &   0.7        &   &  & \\
$\bar{\Omega}$        &   1.765      & 4.4e-04  & 8.0e-06 & 1.0e-06\\
$\bar{M}$             &   0.1804     & 6.7e-04  & 1.7e-04 & 2.5e-05\\
$\bar{R}_{\rm circ}$  &   0.3865     & 4.3e-04  & 1.3e-04 & 2.0e-05\\
$\bar{J}$             &   0.02984    & 2.1e-03  & 4.0e-04 & 6.0e-05\\
$Z_{\rm p}$           &   5.179      & 2.9e-03  & 5.0e-04 & 7.5e-05\\ \hline
$|1-M_{\rm in}/M_{\rm out}|$&        & 1.6e-04  & 9.7e-05 & 4.3e-05\\
$|1-J_{\rm in}/J_{\rm out}|$&        & 4.7e-04  & 1.6e-04 & 6.8e-05\\
          \hline\vspace*{2mm}
         \end{tabular}
         \caption{Results for a homogeneous model with infinite central
	   pressure and $r_{\rm p}/r_{\rm e}=0.7$. For the meaning of the
	   quantities, see Table 1. The virial identities are not defined 
	   for $\bar{p}_{\rm c}=\infty$ since the integrals $I_1,I_2,J_1,J_2$ diverge.}
\end{center}
\end{table}

Another possible endpoint of a sequence of rotating stars in General Relativity
is reached when the pressure diverges at the star's centre. For example, the
sequence of static homogeneous configurations is characterized by this limit. 
Here, the star is spherical ($r_{\rm p}=r_{\rm e}$ and $g=0$), and the 
corresponding gravitational fields are analytically given by the Schwarzschild 
solution which reads in our chosen coordinates (with $r^2=\rho^2+\zeta^2$)
\begin{eqnarray}
{\rm e}^{U'}&=&\frac{3\left[1-M/(2r_{\rm e})\right]}{2\left[1+M/(2r_{\rm
				     e})\right]}
-\frac{1-Mr^2/(2r_{\rm e}^3)}{2+Mr^2/(r_{\rm e}^3)}	\\
W{\rm e}^{-U'}&=&\rho\,\frac{\left[1+M/(2r_{\rm e})\right]^{\,3}}{1+Mr^2/(2r_{\rm
				     e}^3)}		\\
a'&=&0	  
\end{eqnarray}				
inside (i.~e.~for $r<r_{\rm e}$) and
\begin{eqnarray}
{\rm e}^{\nu}&=&\frac{1-M/(2r)}{1+M/(2r)}	\\
W{\rm e}^{-\nu}&=&\rho\,\left[1+M/(2r)\right]^{\,2}		\\
\omega&=&0		  
\end{eqnarray}				
outside the star ($r>r_{\rm e}$). In the limit $Mr_{\rm e}^{-1}\to 1$  the
central value ${\rm e}^{U'_{\rm c}}$ vanishes which corresponds to $p_{\rm c}\to \infty$
since the surface potential $V_0=U'(r=r_{\rm e})$ remains finite.

A rotating configuration with an infinite central pressure is characterized by
an ergo-region that extends in the inside up to the coordinate origin, see Fig. 3.
Hence, at this point the space-time violates the requirement of
stationarity\footnote{A locally stationary, axisymmetric spacetime requires the
existence of a timelike linear combination of the two Killing vectors
corresponding to stationarity and axisymmetry. The latter one vanishes on the
symmetry axis.} and therefore
some irregular behaviour of the gravitational potentials arises here, which, in the
slow rotation limit, has been studied by Chandrasekhar \& Miller (\cite{CM74}). 
Consequently, we again expect a failure of the AKM-method. But as in the case 
when the mass-shedding limit occurs, we are still able to obtain slowly
converging numerical solutions, see Table 7. It is however necessary (i) to
modify the Chebyshev representation of the interior gravitational potentials 
and (ii) to introduce a slightly different coordinate mapping of the interior
region. In particular, this is done by writing
\begin{eqnarray}
   {\rm e}^{U'}   &=& {\rm e}^{V_0}+(s-1)\tilde{H}_{U'}(s,t)\\[3mm]
   a'{\rm e}^{U'} &=& \rho^2\left[\tilde{a}'_{\rm B}(t)+(s-1)\tilde{H}_{a'}(s,t)\right]\\[3mm]
   W^{\rm (int)}{\rm e}^{-U'}  &=& \rho\;{\rm e}^{-V_0}\;
   \left[1+\tilde{W}_{\rm B}(t)+(s-1)\tilde{H}_{W,\rm int}(s,t)\right]
\end{eqnarray}
and using again the transformation (\ref{Transf_Int1}). The reformulation
of the interior Chebyshev expansions is motivated by the Schwarzschild solution
given above. We learn from here that the vanishing of ${\rm e}^{U'_{\rm c}}$ 
coincides with that of the central value of $W/\rho$. 
Moreover, we note that the combination 
\begin{eqnarray}\rho^{-2}g_{\varphi\varphi}=(\rho^{-1}W{\rm e}^{-U'})^2-(\rho^{-1}a'{\rm
e}^{U'})^2\end{eqnarray}
remains positive (and finite) at the origin when ${\rm e}^{U'_{\rm c}}\to 0$. So the
above reformulation ensures particular dependencies of the metric functions at the
origin when ${\rm e}^{U'_{\rm c}}\to 0$.
 
The use of the transformation (\ref{Transf_Int1}) allows one to lay the coordinate
mesh more densely in the vicinity of the origin. This helps to take the
singular behaviour in higher derivatives of the functions $\tilde{H}_{U'},\tilde{H}_{a'}$ 
and $\tilde{H}_{W,\rm int}$ into account and thus provides a better convergence.
For the example given in Table 7 and Fig. 3, we used $c_{\rm s}=0.65$. 
In the approximation scheme we prescribed the parameters 
(${\rm e}^{U'_{\rm c}},r_{\rm p}/r_{\rm e}$) and finally pushed ${\rm
e}^{U'_{\rm c}}$ to zero.

\subsubsection{Highly flattened stars}
\label{Flat}
\begin{table}
\begin{center}
   \begin{tabular}{lllll}\hline
     \multicolumn{2}{c}{}&\multicolumn{1}{c}{m=10}&
     \multicolumn{1}{c}{m=16}&\multicolumn{1}{c}{m=22}\\ \hline
$\bar{p}_{\rm c}$     &   0.002         &   &  & \\
$r_{\rm p}/r_{\rm e}$ &   0.2           &   &  & \\
$\bar{\Omega}$        &   1.089864e-00  & 2.8e-06  & 9.4e-08 & 6.8e-10\\
$\bar{M}$             &   8.371248e-04  & 4.0e-05  & 1.8e-06 & 2.3e-08\\
$\bar{R}_{\rm circ}$  &   1.027320e-01  & 5.6e-06  & 2.2e-07 & 2.5e-09\\
$\bar{J}$             &   3.703716e-06  & 1.7e-04  & 2.7e-06 & 3.7e-08\\
$Z_{\rm p}$           &   1.592971e-02  & 3.3e-05  & 1.3e-06 & 1.6e-08\\ \hline
$GRV2$&                                 & 4.2e-06  & 4.7e-08 & 6.3e-10 \\ 
$GRV3$&                                 & 6.0e-06  & 6.2e-08 & 8.5e-10 \\
$|1-M_{\rm in}/M_{\rm out}|$&           & 2.2e-06  & 5.4e-08 & 4.2e-10\\
$|1-J_{\rm in}/J_{\rm out}|$&           & 1.0e-04  & 3.8e-07 & 1.8e-09\\
          \hline\vspace*{2mm}
         \end{tabular}
         \caption{Results for a homogeneous model with $\bar{p}_{\rm c}=0.002$ 
	     and $r_{\rm p}/r_{\rm e}=0.2$. For the meaning of the quantities,
	     see Table 1.}
\end{center}
\end{table}
\begin{figure}
\unitlength1cm
\hspace*{-1cm}
\begin{center}
\psfrag{r}[r][r]{{$\rho$}}
\psfrag{q}[r][r]{{$\zeta$}}
\epsfig{file=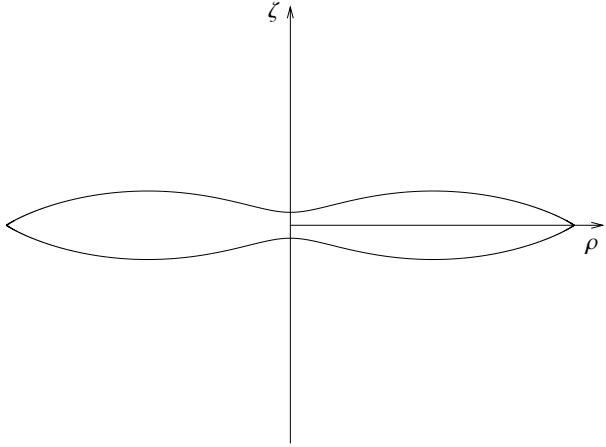,scale=0.7}
\end{center}
\caption{Meridional cross-section of a highly distorted 
homogeneous configuration 
at the mass shedding limit, specified in Table 9 
(with the axes scaled identically).}
\end{figure}
\begin{table}
\begin{center}
   \begin{tabular}{lllll}\hline
     \multicolumn{2}{c}{}&\multicolumn{1}{c}{m=10}&
     \multicolumn{1}{c}{m=16}&\multicolumn{1}{c}{m=22}\\ \hline
$\bar{p}_{\rm c}$     &   0.003      &   &  & \\
$r_{\rm p}/r_{\rm e}$ &   0.04534    & 1.1e-02  & 1.3e-03 & 1.3e-04\\
$\bar{\Omega}$        &   0.9883     & 1.5e-03  & 3.9e-05 & 1.9e-06\\
$\bar{M}$             &   0.03900    & 6.9e-03  & 9.8e-04 & 5.5e-05\\
$\bar{R}_{\rm circ}$  &   0.4005     & 2.5e-03  & 6.9e-05 & 2.3e-05\\
$\bar{J}$             &   0.002717   & 1.2e-02  & 1.7e-03 & 9.5e-05\\
$Z_{\rm p}$           &   0.2388     & 4.8e-03  & 7.8e-04 & 4.4e-05\\ \hline
$GRV2$&                              & 6.2e-05  & 2.3e-05 & 5.9e-06 \\ 
$GRV3$&                              & 1.3e-04  & 3.0e-05 & 7.3e-06 \\
$|1-M_{\rm in}/M_{\rm out}|$&        & 1.5e-04  & 2.1e-05 & 4.6e-06\\
$|1-J_{\rm in}/J_{\rm out}|$&        & 2.0e-03  & 1.3e-05 & 8.8e-06\\
          \hline\vspace*{2mm}
         \end{tabular}
         \caption{Results for a highly distorted 
                  homogeneous configuration at the mass shedding limit
			with $\bar{p}_{\rm c}=0.003$.
	     For the meaning of the quantities, see Table 1.}
\end{center}
\end{table}
As with the situations above, the coordinate transformations 
(\ref{Transf_Int},\ref{Transf_Ext}) need to be modified if one wants to
calculate models of strongly distorted stars (such as the examples given 
in table 2 of Ansorg et al. \cite{AKM02}). When
using (\ref{Transf_Int},\ref{Transf_Ext}), then each
curve $s={\rm const}$ represents an image which is similar to the star's
boundary. This leads for distorted stars 
to a non-uniform partition of the domains by the coordinate net of
($s,t$)-variables. Moreover, the oblateness of the configurations suggests
adapting the coordinates $s$ and $t$ for the exterior domain to resemble oblate 
spheroidal coordinates. A possible mapping which takes these considerations into
account is given by [$(s,t)\in I^2$]:
\begin{eqnarray} \label{Transf_Int2}
\rho^2&=&r_{\rm e}^2t\,\tau(t)\,,\\
\zeta^2&=&s(1-t)[r_{\rm p}^2+tg(t)]
\end{eqnarray}
for the interior region and
\begin{eqnarray} \label{Transf_Ext2}
\rho^2&=&t[r_{\rm e}^2-r_{\rm p}^2+\xi^2(s)][1-s+s\tau(t)],\\
\zeta^2&=&(1-t)[\xi^2(s)+tg(t)] 
\end{eqnarray}
exterior to the star with
\begin{eqnarray}\tau(t)=\frac{1}{1-c_{\rm t}(1-t)}\,,\quad\xi(s)=r_{\rm p}+\frac{c_{\rm
s}}{s}(1-s).\end{eqnarray}
For the examples (b) and (c) in Ansorg et al. (\cite{AKM02})  we took
$c_{\rm s}=0.07$ and $c_{\rm t}=-0.2$ whereas for the example (a) in Ansorg et al. (\cite{AKM02})
as well as for that presented in Table 8 the values $c_{\rm s}=0.2$ and 
$c_{\rm t}=0$ were chosen. Note that the
prescribed parameter pair, ($p_{\rm c},r_{\rm p}/r_{\rm e}$), is the same for
the latter two configurations. All other physical quantities are also very similar
for these models, including the appearance of the corresponding cross sections. 
Here, the high accuracy is needed in order to distinguish between 
these two nearby configurations. 

A final example (see Table 9 and Figure 4) exhibits that even in the highly
flattened regime,
configurations at the mass shedding limit can be calculated (here 
$c_{\rm s}=0.2 $ and $c_{\rm t}=0 $). This particular
model is close to the configuration (a) in Ansorg et al. (\cite{AKM03a}),
which marks the transition body
between spheroidal and toroidal configurations at the mass shedding limit. A more
detailed investigation of highly flattened homogeneous bodies in General
Relativity will be published elsewhere.

\begin{acknowledgement}
      The authors would like to thank David Petroff and 
	Nik Stergioulas for many valuable discussions.
      This work was supported by the German
      \emph{Deut\-sche For\-schungs\-ge\-mein\-schaft\/} (DFG-project
      ME~1820/1-3).
\end{acknowledgement}

\end{document}